# Unravelling the Use of Digital Twins to Assist Decision- and Policy-Making in Smart Cities

Lucy Temple,[1] Gabriela Viale Pereira,[1] Lukas Daniel Klausner[2]

[1] University for Continuing Education Krems, Department for E-Governance and Administration, Krems, Austria
lucy.temple@donau-uni.ac.at, gabriela.viale-pereira@donau-uni.ac.at
[2] St. Pölten University of Applied Sciences, Department of Computer Science and Security, St. Pölten, Austria
mail@l17r.eu

This short paper represents a systematic literature review that sets the basis for the future development of a framework for digital twin-based decision support in the public sector, specifically for the smart city domain. The final aim of the research is to model context-specific digital twins for aiding the decision-making processes in smart cities and devise methods for defining the policy agenda. Overall, this short paper provides a foundation, based on the main concepts from existing literature, for further research in the role and applications of urban digital twins to assist decision- and policy-making in smart cities. The existing literature analyses common applications of digital twins in smart city development with a focus on supporting decision- and policy-making. Future work will centre on developing a digital-twin-based sustainable smart city and defining different scenarios concerning challenges of good governance, especially so-called wicked problems, in smaller-scale urban and non-urban contexts.

# 1   Introduction

The importance of an information systems (IS) focus on smart-city-related studies is highlighted by Ismagilova et al. (2019), especially regarding the implementation and use of IS to design, develop and plan smart cities. Following the approach by Ismagilova et al. (2019, p. 90): Smart cities use an IS-centric approach to the intelligent use of information and communication technologies (ICT) within an interactive infrastructure to provide advanced and innovative services to its citizens, impacting quality of life and sustainable management of natural resources. Emerging technologies that serve to support human decision-making are transforming existing government arrangements and are very promising for decision support in the context of smart cities, both in the policy process and in operative decision-making (König & Wenzelburger, 2021). Digital twins are considered an emerging technology in smart city research (Hämäläinen, 2021). Although digital twin technology originally emerged as a tool for designing virtual replicas in manufacturing, it has now evolved into a concept that can be applied to different domains, from engineering through automobile manufacturing to energy supply systems (White et al., 2021), and existing research has shown a growing interest in their application to facilitate decision-making for urban planning and policy decisions alike (Lei et al., 2023; White et al., 2021). Digital twins of cities can assist policymakers in the process of making strategic long-term decisions regarding urban planning (Lohman et al., 2023), as well as supporting the planning and management of cities in urban and peri-urban areas (Lei et al., 2023). According to Lohman et al. (2023), digital twins connect data, analytics and visualisation, enabling policymakers to simulate what-if scenarios in a scenario-based analysis. While "analytics" means the growth of data that leads to new technological and scientific developments, "policy analytics" consists in the use of analytics to support public policy decision-making

(Daniell et al., 2016, p. 7). Bringing this to a public sector perspective, digital technologies can be used to support different government functions, e. g. legislative (for policymaking), management (for policy advice and administrative tasks) and service delivery through employees and service providers. In the case of local government, according to Clement and Crutzen (2021), a local political system and objectives are supported by the so-called smart city policy domain, which defines the smart city policy agenda as the list of topics to address local problems that are influenced by the local context. The agenda is set when a given problem aligns with an appropriate smart city solution and is also aligned with policy priorities at the local level. Yossef Ravid and Aharon-Gutman (2023) show the potential of digital twins to also incorporate social aspects into the decision-making process. We use this idea as the basis for analysing the role and applications of digital twins to support urban planning, city management, smart city initiatives and policy in smart cities. When designing urban planning policies, such as deciding where to create a new park, data retrieved from the physical smart city such as air pollution, noise pollution, traffic flow and sunlight can be used to simulate various park placement options in the digital twin. By using smart sensors, the flow of people in the area can also be simulated to help decide where benches, fitness equipment, paths etc. need to be planned (Ramu et al., 2022; White et al., 2021).

The main objective of this ongoing research is to review the existing literature on the intersection of digital twins and smart cities with a focus on decision and policy-making support and to answer the research question: "What are the existing applications of digital twins for smart cities for aiding decision-and policy-making?" This work in progress forms part of the baseline research conducted within the project Smart Cities and Digital Twins in Lower Austria (SCiNDTiLA), which explores how the concept of smart cities can be transferred to smaller-scale urban and non-urban contexts and how the use of digital twins and algorithms can aid decision support and policy-making towards developing smart sustainable solutions. A variety of use cases will be defined within Lower Austria and digital twins will be simulated to provide scenarios to be used for local decision-making.

## 2    Systematic Literature Review Methodology

This short paper analyses the existing literature at the intersection of digital twins and smart cities with a focus on the application of this technology as a tool to support decision- and policy-making for city managers. For the systematic literature review, we explored the existing literature on digital twins and smart cities, using the following search in the Scopus database: "⟨title OR abstract OR author-specified keywords: "smart cit*" AND "digital twin*" AND ⟨"decision support" OR "decision-making"⟩⟩ AND ⟨publication year: >2013⟩". We found 94 relevant records; we then scanned the titles and abstracts and selected 47 papers which had a clear focus on supporting decision-making. After a full-text screening of these papers, the preliminary analysis included 25 papers. The full analysis of the selected articles focused on the variety of uses of digital twins in smart cities, the tools involved to gather the relevant data needed, the simulations these can produce and the challenges and enablers identified. The following section presents the relevant findings from the literature and is subdivided into two subsections: data and tools as well as challenges and enablers. These themes were derived by grouping the findings of the selected articles, allowing us to synthesise and interpret the literature to provide a comprehensive understanding of the current state of research in the field.

## 3    Literature Analysis

A digital twin, as defined in literature (White et al., 2021), is a virtual representation of a physical process, person, place, system or device, initially developed to enhance manufacturing processes through precise simulations featuring highly accurate models of individual components. The overarching aim of digital twins is to simulate the behaviour of the targeted object and enable real-time decision-making based on reasonable predictions (Shi et al., 2023). This entails capturing real-time characteristics and status to facilitate proactive actuation orders. A more comprehensive definition, as proposed by Zhou et al. (2022), underscores the creation of a virtual representation for a dynamic physical object or system, spanning multiple life cycle stages and facilitating decision-making through the application of data analysis methods.

### 3.1    Data and Tools

With the rapid digitisation of cities, there is more and more data available for use and analysis. However, often the raw data are insignificant without being embedded in the right context. Moreover, the amount of data created exceeds human capabilities for simple analysis and prediction. These data can be interlinked and placed in simulated environments, and with the right tools, a Smart City Digital Twin (SCDT) can be modelled. As cities are complex systems, digital twins of smart cities should focus on the interdependencies between networks and city infrastructure to be able to accurately represent the physical object in its specific usage context (Mohammadi et al., 2020). The variety of components within a digital twin work together to create a virtual representation or urban portrait of the physical city. This allows for real-time monitoring, but also analysis and optimisation. The availability of spatial and non-spatial data alongside evolving simulation technologies allows for the digital representation of such intricate entities as cities (Jeddoub et al., 2023).

Relevant types of data and technologies used in the context of a physical smart city include the internet of things (IoT), geographic information systems (GIS), building information modelling (BIM), natural language processing (NLP) and machine learning (ML) (Shi et al., 2023; White et al., 2021; Yaqoob et al., 2023). Data used for simulating digital twins of physical counterparts must be of high quality, accurate, valid, complete and consistent. This will allow producing more valuable and reliable scenarios for policy makers to base their decisions on (Li & Tan, 2023). A SCDT is expected to be used as a means to achieve evidence-based decisions, not as an end, leading to better outcomes and more informed policy processes (Wan et al., 2019). Petrova-Antonova and Ilieva (2019) believe such simulations allow for discussion between stakeholders to select the best outcome and get insights regarding the possible effects after deployment.

The design and implementation of a SCDT can provide a broader vision for future planning and city improvements by using computer modelling and information technologies to diagnose and map aspects present in the physical world, allowing for high efficiency and intelligent decision-making (Lyu et al., 2022). In order to do so, the SCDT requires high-quality, long-term data for decision-making (Ramu et al., 2022). Weil et al. (2023) state that all digital infrastructure and sensors in a smart city can be used more efficiently for decision-making processes thanks to the advances in SCDT, allowing for simulation models and predictions. By using real-time data to simulate the system's performance, it allows for real-time monitoring, to forecast and optimise the physical counterpart, but also for analysis and streamlining to allow for faster and more accurate predictions, better decision-making, and quicker response time (Li & Tan,

2023; Wang et al., 2023). The SCDT can be used as a decision support tool, but the agency of the actors is still necessary for decision-making. The technology supports the decision process by linking the actors with the resources and information (West et al., 2021).

As digital twins are able to store and process more historical and real-time data, they become capable of predicting and forecasting variations, visualising what-if scenarios that can help city decision-makers make proactive decisions (such as for disaster prevention) (Li & Tan, 2023; Mavrokapnidis et al., 2021). SCDT can enable an overview of heterogeneous data instead of isolated interpretations of specific datasets (Raes et al., 2021). This holistic analysis and visualisation approach allows for policymakers to process and use heterogeneous city data, integrating their domain expertise with the information provided by the digital twin and thus supporting integrated management decisions, macro decision-making and evaluation (Lyu et al., 2022; Mohammadi et al., 2020).

### 3.2  Challenges and Enablers

Ramu et al. (2022) state that digital twins are still in a very early stage of usage for smart city applications; the main reason for this is the lack of trust and privacy issues of sharing sensitive data. SCDT are commonly used in the planning, design and development of a city, its core policy agenda and for both short- and long-term planning (Mendula et al., 2022; Weil et al., 2023). Weil et al. (2023) highlight the importance of cooperation between decision-makers in order for the SCDT to become a useful tool – data-sharing, joint planning, political support and dialogue are needed. Despite the increased popularity of SCDT, there still exists a series of bottlenecks for their implementation. First, it is crucial to have consistency between the physical and the virtual representation. There is a lack of centralised, unified and defined frameworks regarding how to make the data connections between the digital and the physical (Lei et al., 2023; Ma et al., 2024; Raes et al., 2021). Challenges identified in the literature range from technical (interoperability and semantic standards) to non-technical (the need to be purposeful, trustworthy and functional, or the lack of business models) (Lei et al., 2023; Weil et al., 2023). In order to be able to represent the physical city and support decision- and policy-making, a digital twin must be able to break through data silos while at the same time providing secure information (Shi et al., 2023; Yaqoob et al., 2023). Moreover, the data quality is key for presenting policymakers with accurate data and reliable results and measurements (Weil et al., 2023). Early SCDT platforms relied on IoT implementations throughout smart cities, but these lacked the scale and suitability to be useful in making policy decisions (Raes et al., 2021). Many authors still claim that digital twins are lacking the ability to integrate the socio-economic and human dynamics of cities (Mohammadi & Taylor, 2019). White et al. (2021) propose stakeholder and citizen involvement in the SCDT. Citizens identify problem areas within the city and give feedback to proposed policies; this information is fed back into the SCDT to create additional data. Not all SCDT are designed for citizen engagement, but when users are able to participate, comment and request changes by assessing the virtual environment, they are able to single out existing issues. To adequately include citizens' voice in decision-making, a SCDT needs to receive human input for allowing human participation in decision-making (Abdeen et al., 2023; Ramu et al., 2022).

It is important for SCDT to consider changes in organisational culture, processes and structures allowing for a clear digital data flow between the cities' information systems and the entities within the cities (Hämäläinen, 2021).The use of a digital twin allows for simulating a series of what-if scenarios, which can be immensely helpful in such contexts. Time and computational capabilities remain key challenges for SCDT; decision support scenarios ideally need to be evaluated in a series of workshops or focus groups, and to that end, computational times need to be manageable, with synchronicity and the rate at which data can flow key for the practical usefulness of SCDT (Jeddoub et al., 2023; Lohman et al., 2023; Wang et al., 2023; Weil et al., 2023).

# 4      Conclusions and Future Work

This literature review analyses common applications of digital twins in smart city development for assisting decision support. This research in progress highlights the significance of information systems in smart-city-related studies and the potential of digital twin technology to assist in designing, planning and developing smart cities, specifically how digital-twin-based decision support can aid policymakers to make better decisions. The empirical research will be conducted in Lower Austria, which is characterised by the growing number of policies that have been developed to support "digitalisation of the public sector" and, especially, "smart initiatives" at the regional and local level according to the Digitalization Strategy of Lower Austria. Future work includes developing a SCDT and defining different scenarios concerning challenges of good governance in smaller-scale urban and non-urban contexts as well as implementing this proof of concept in an exemplary region and using the insight gained to draft a roadmap, highlighting methodologies, guidelines and policy recommendations on how smart and sustainable solutions in cities and regions shape inhabitants' perception of local governance.

**Acknowledgements**

This research was funded by the Gesellschaft für Forschungsförderung Niederösterreich (GFF NÖ) project GLF21-2-010 "Smart Cities and Digital Twins in Lower Austria". The financial support by the Gesellschaft für Forschungsförderung Niederösterreich is gratefully acknowledged.